\documentclass[letterpaper, 10 pt, conference]{ieeeconf}  

\IEEEoverridecommandlockouts                              

\usepackage[T1]{fontenc}	
\usepackage[utf8]{inputenc}
\usepackage{graphics} 
\usepackage{epsfig} 
\usepackage{amsmath,mathtools,amsfonts,amsthm} 
\usepackage{amssymb}
\usepackage{accents}
\usepackage{verbatim}
\usepackage{enumitem,kantlipsum}
\usepackage{color}
\usepackage{cite}

\newtheorem{thm}{Theorem}

\newtheorem{rem}{Remark}
\newtheorem{ass}{Assumption}

\title{\LARGE \bf
Robust Trajectory Tracking for Underactuated Quadrotors with Prescribed Performance*
}

\author{Dženan Lapandić$^{1}$, Christos K. Verginis$^{2}$, Dimos V. Dimarogonas$^{1}$ and Bo Wahlberg$^{1}$
\thanks{*This work was supported by the Wallenberg AI, Autonomous Systems and Software Program (WASP), the Swedish Research Council, Knut and Alice Wallenberg Foundation (KAW) and the SSF COIN project.}
\thanks{$^{1}$Dženan Lapandić, Dimos V. Dimarogonas and Bo Wahlberg are with  Division of Decision and Control Systems, KTH Royal Institute of Technology, Stockholm, Sweden.   {\tt\small lapandic,dimos,bo@kth.se}}%
\thanks{$^{2}$Christos K. Verginis is with Division of Signals and Systems, Department of Electrical Engineering, Uppsala University,
Uppsala, Sweden. {\tt\small christos.verginis@angstrom.uu.se
}
         }%
}

\begin{document}

\maketitle
\thispagestyle{empty}
\pagestyle{empty}

\begin{abstract}

We propose a control protocol based on the prescribed performance control (PPC) methodology for a quadrotor unmanned aerial vehicle (UAV). Quadrotor systems belong to the class of underactuated systems for which the original PPC methodology cannot be directly applied. 
We introduce the necessary design modifications to stabilize the considered system with prescribed performance. The proposed control protocol does not use any information of dynamic model parameters or exogenous disturbances. Furthermore, the stability analysis guarantees that the tracking errors remain inside of designer-specified time-varying functions, achieving prescribed performance independent from the control gains' selection. Finally, simulation results verify the theoretical results.
\end{abstract}

\section{INTRODUCTION}
Coordination and control of unmanned aerial vehicles (UAV) has drawn considerable attention in recent years. Despite its numerous applications such as exploration, delivery, patrolling, search and rescue missions, there are still unresolved challenges connected with this control problem. UAV systems are highly nonlinear, underactuated and model parameters may vary during the flight. This makes the control design even more demanding, especially in scenarios when UAVs need to meet performance and safety specifications.  Such specifications are vital in landing scenarios on other unmanned ground vehicles (UGV). Landing scenarios and agents coordination have been explored in \cite{vlantis2015quadrotor,persson2019model,paris2020dynamic,lapandic2021aperiodic}.

There already exists an extensive amount of works in literature concerning stabilization and trajectory tracking control of quadrotors. The early works consider proportional-integral-differential (PID) controllers  \cite{bouabdallah2004pid,luukkonen2011modelling,li2011dynamic}, which is designed on simplified model excluding cross-coupling in attitude dynamics and has limited performance in the presence of strong perturbations, and linear-quadratic regulator (LQR) \cite{argentim2013pid,tran2015quadrotor,foehn2018onboard}, whose limitations stem from linearization and requirement of model knowledge. 
Advanced control methods such as backstepping \cite{madani2006backstepping,aboudonia2017active} and sliding-mode control \cite{bouabdallah2005backstepping} deliver satisfactory tracking performance but still require model knowledge. Sliding-mode control is known for introducing chattering effect and improvements are made using adjusted boundary-layer sliding control \cite{runcharoon2013sliding,lopez2018robust}. Adaptive control based on backstepping is derived in \cite{huang2010adaptive} and $\mathcal{L}_1$~adaptive control is used for aggressive flight maneuvers in \cite{michini2009l1}. Flatness-based control is employed in \cite{formentin2011flatness,sreenath2013geometric}, and $\mathcal{H}_\infty$~controller in \cite{raffo2009underactuated}.
With the increase of available computational power in embedded devices, Model predictive control (MPC) became very popular due its ability to handle state and input constraints and optimize the trajectory online \cite{kamel2017linear,nguyen2021model}. For handling the disturbances, various wind estimation model-based and neural network methods have been applied \cite{sydney2013dynamic,tagliabue2020touch}.

However, most of these works focus on model-based approaches or the stability can only be shown around linearized equilibrium points. Furthermore, a significant property that lacks from the related
literature on quadrotor control is tracking/stabilization with predefined transient and steady-state specifications, such
as overshoot, convergence speed or steady state error. Such specifications can encode time and safety constraints, which are crucial when it comes to physical autonomous systems, and especially UAVs. 

In this paper, we develop a modified Prescribed Performance Control (PPC) protocol, which traditionally deals with model uncertainties and transient- and steady-state constraints \cite{BECHLIOULIS20141217}, to solve the trajectory-tracking control problem for quadrotors with prescribed performance.

Our main contribution is in the extension of the original PPC algorithm to account for the underactuated quadrotor system. At the same time, the proposed control protocol does not use any information on the model parameters and is robust to unknown exogenous disturbances without employing approximation or observer-based schemes. Similarly to the original PPC methodology, the tracking errors evolve within predefined user-specified functions of time, achieving prescribed transient and steady-state performance that is independent from the selection of the control gains.  
 
It should be noted that PPC has been recently used to control quadrotors. In \cite{chang2017adaptive,xu2020adaptive} authors use PPC for the attitude subsystem, thus avoiding the underactuated part of the system. Other works  \cite{hua2018adaptive,jiang2020composite,sasaki2020disturbance,shen2021prescribed} focus on the complete system but use neural network approximations, partial knowledge of dynamic parameters, observers for disturbance estimates and exploit these information in the controller. On the contrary, the proposed controller does not use any information on the dynamic parameters or external disturbances. In \cite{verginis2022robust}, the authors proposed a similar control design using PPC for an underactuated 3-DOF helicopter. Such a system is, however, significantly different than the one studied in this paper and hence the respective control design is not applicable.

The rest of the paper is organized as follows. First, we provide necessary preliminaries in Section~\ref{sec:preliminaries} and problem formulation in Section~\ref{sec:prob_statement}. Then, we present the proposed control design and stability analysis in Section~\ref{sec:main_results}, and Section~\ref{sec:simulation_results} illustrates the effectiveness of the controller. Finally, Section~\ref{sec:conclusion} concludes the paper. 

\section{PRELIMINARIES}\label{sec:preliminaries}

In this section, we present the basic framework of prescribed performance control which is originally introduced in \cite{BECHLIOULIS20141217}. The main idea is to ensure the convergence of the tracking errors $e(t)$ within a predefined set at a specified rate. This is accomplished by enforcing the error to stay within a region bounded by a certain smooth and bounded function of time, i.e.,
\begin{equation}
    -\rho(t) < e(t) < \rho(t), \quad \forall t \geq 0,
\end{equation}
where $\rho(t)$ is the performance prescribed function that can be defined as
\begin{equation}
    \rho(t) = (\rho_{0} - \rho_{\infty})e^{-l t} + \rho_{\infty}, \quad \forall t \geq 0, \label{eq:performance_fcn}
\end{equation}
with positive chosen constants $\rho_{0},\rho_{\infty},l>0$. 
\begin{figure}[h!]
      \centering
      \includegraphics[width=0.8\linewidth]{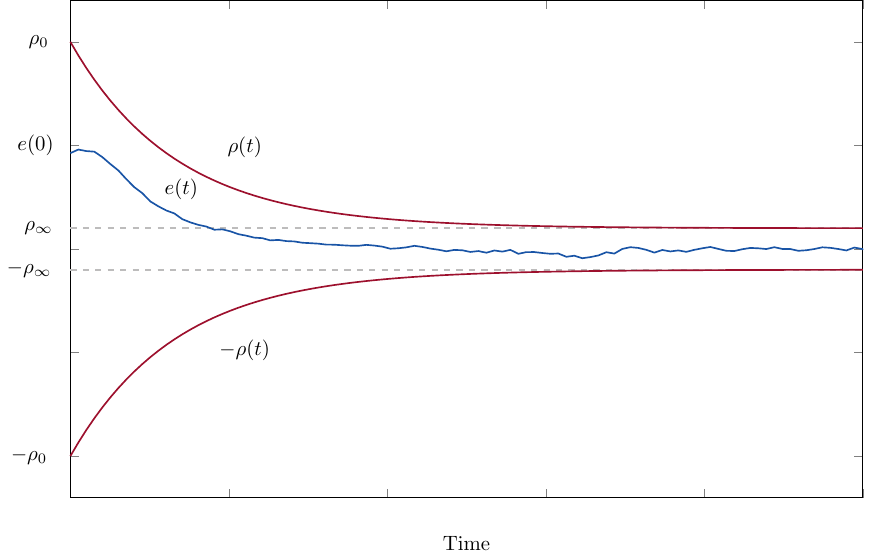}
      \vspace{-0.3cm}
      \caption{The error evolves inside of the prescribed performance funnel. }
      \label{fig:sketch}
\end{figure}

Practically, $\rho_{0}$ is selected such that the error starts inside of the prescribed funnel, i.e. $\rho_0 > |e(0)|$, $\rho_{\infty}:=\lim_{t\rightarrow\infty}\rho(t)>0$ represents the upper bound on the steady-state error and $l$ is the lower bound on the convergence rate of the error. Therefore, appropriate choice of the discussed parameters determines the transient and steady-state performance of the error $e(t)$, as depicted on Fig.~\ref{fig:sketch}. Furthermore, let the normalized errors $\xi(t)$ be defined as
\begin{equation*}
    \xi(t) = \rho(t)^{-1} e(t).
\end{equation*}
The important point in designing PPC is a transformation of the normalized error $\xi(t)$ with a strictly increasing, bijective function $\mathrm{T}: (-1,1) \rightarrow (-\infty,\infty)$
\begin{equation}
    \mathrm{T}(\xi(t)) =  \textup{atanh}(\xi(t)) =  \frac{1}{2} \ln{\frac{1+\xi(t)}{1-\xi(t)}}.\label{eq:transformation}
\end{equation}
The derivative of this transformation is $\frac{\mathrm{d} \mathrm{T}(\xi)}{\mathrm{d} \xi} = \frac{1}{1-\xi^2}.$
%
For a vector $\boldsymbol{\xi} = [\xi_1,\dots,\xi_n]^T \in \mathbb{R}^n$, we define 
\begin{equation}\label{eq:transf_vec}
    \mathrm{T}(\boldsymbol{\xi}) = 
    \frac{1}{2}
    \begin{bmatrix}\ln{\frac{1+\xi_1}{1-\xi_1}} & \dots &
    \ln{\frac{1+\xi_n}{1-\xi_n}} \end{bmatrix}^T
\end{equation}
\begin{equation} \label{eq:r_matrix_def}
    \frac{\mathrm{d} \mathrm{T}(\boldsymbol{\xi})}{\mathrm{d} \boldsymbol{\xi}}  = \textup{diag}\left\{ \left[\frac{1}{1-\xi_i^2} \right]_{i \in \{1,\dots,n\} } \right\}
\end{equation}

\section{PROBLEM STATEMENT}\label{sec:prob_statement}
We consider the quadrotor UAV model:
\begin{subequations}
\begin{align}
    \dot{\boldsymbol{p}} &= \boldsymbol{v},\\
    \dot{\boldsymbol{v}} &= \frac{1}{m}\left (\boldsymbol{R}_{IB}(\boldsymbol{\eta})  \boldsymbol{F}_{T} +  \boldsymbol{F}_{d}( \boldsymbol{\chi}, \dot{\boldsymbol{\chi}},t) \right ) - \boldsymbol{g},\label{eq:1b} \\
    \dot{\boldsymbol{\eta}} &= \boldsymbol{R}_{T}(\boldsymbol{\eta}) \boldsymbol{\omega},\\
    \boldsymbol{I}(\boldsymbol{\eta}) \dot{\boldsymbol{\omega}} &= - \boldsymbol{\omega}\times \boldsymbol{I}(\boldsymbol{\eta}) \boldsymbol{\omega} + \boldsymbol{\tau} + \boldsymbol{\tau}_d( \boldsymbol{\chi}, \dot{\boldsymbol{\chi}},t)
\end{align}\label{eq:system}
\end{subequations}
where $\boldsymbol{\chi} = [ \boldsymbol{p}^T, \boldsymbol{v}^T, \boldsymbol{\eta}^T, \boldsymbol{\omega}^T]^T$, $\boldsymbol{p} \in \mathbb{R}^3$ is the position in the inertial frame, $\boldsymbol{v} \in \mathbb{R}^3$ is the linear velocity, $\boldsymbol{\eta} = [\phi,\theta,\psi]^T \in \mathbb{T} = (-\frac{\pi}{2},\frac{\pi}{2})\times(-\frac{\pi}{2},\frac{\pi}{2})\times (-\pi,\pi)$ is the vector of Euler angles representing the attitude (roll, pitch, yaw angles), and $\boldsymbol{\omega} = [\omega_\phi,\omega_\theta,\omega_\psi]^T$ is the angular velocity, expressed in the inertial frame; 
$\boldsymbol{F}_{T} = [0,0,F_z]^T$ is the controlled thrust, and $\boldsymbol{\tau}$ is the inertial-frame controlled torque;
$\boldsymbol{R}_{IB}: \mathbb{T} \to SO(3)$ is the rotation matrix from the body to the inertial frame, and $SO(3)$ is the special orthonormal group $SO(n) = \{ \boldsymbol{R} \in \mathbb{R}^{n\times n} : \boldsymbol{R}\boldsymbol{R}^T = I_n, \det \boldsymbol{R} = 1  \}$ in 3D and $I_n \in \mathbb{R}^{n\times n}$ is the identity matrix; $\boldsymbol{R}_{IB}$ is
 the product of three consecutive rotations for angles $\psi,\theta,\phi$ around $z,y,x$ axes, respectively, i.e., $\boldsymbol{R}_{IB}(\boldsymbol{\eta})  = \boldsymbol{R}_{z,\psi}(\psi)\boldsymbol{R}_{y,\theta}(\theta)\boldsymbol{R}_{x,\phi}(\phi)$ with $\boldsymbol{R}_{z,\psi}:(-\pi,\pi) \to SO(3)$, $\boldsymbol{R}_{y,\theta}:(-\frac{\pi}{2},\frac{\pi}{2}) \to SO(3)$, $\boldsymbol{R}_{x,\phi}:(-\frac{\pi}{2},\frac{\pi}{2})\to SO(3)$ the respective rotation matrices.
 Furthermore, 
 $\boldsymbol{R}_{T}:\mathbb{T}\to\mathbb{R}^{3\times 3}$ is the mapping from the angular velocity to the time derivatives of the Euler angles
\begin{equation*}
     \boldsymbol{R}_{T}(\boldsymbol{\eta}) = \begin{bmatrix} 
     \frac{ c_\psi}{ c_\theta} & \frac{ s_\psi}{ c_\theta} & 0 \\
     - s_\psi &  c_\psi & 0 \\
      c_\psi  t_\theta &  s_\psi  t_ \theta & 1
           \end{bmatrix}.
\end{equation*}
where we adopt the shorthand notation for trigonometric functions, i.e., $s_\psi = \sin\psi, c_\psi = \cos \psi, t_\theta = \tan \theta$. Note that $\boldsymbol{R}_{T}$ is well-defined for $\theta \in (-\frac{\pi}{2},\frac{\pi}{2})$, which we assume in this paper. The functions $\boldsymbol{F}_{d}(\boldsymbol{\chi},\dot{\boldsymbol{\chi}},t) = [\boldsymbol{F}_{d,xy}^T, \boldsymbol{F}_{d,z}]^T$, $\boldsymbol{\tau}_d:=\boldsymbol{\tau}_d(\boldsymbol{\chi},\dot{\boldsymbol{\chi}},t)$ represent unmodelled aerodynamic forces and moments like drag, hub forces or ground and gyroscopic effects, and exogenous disturbance.
The two functions are continuous in $\boldsymbol{\chi}$ and $\dot{\boldsymbol{\chi}}$, uniformly bounded in $t$. The term $\boldsymbol{g} = [0,0,g]^T\in \mathbb{R}^3$ corresponds to the constant gravity vector. Finally, $m\in \mathbb{R}$ and $\boldsymbol{I}:\mathbb{T}\to \mathbb{R}^{3 \times 3}$ are the mass and positive definite inertia matrix of the UAV, 
also considered \textit{unknown}.

In this paper, we consider the tracking control problem of 
given time-varying reference trajectories $\boldsymbol{p}_r = [p_{x,r},p_{y,r},p_{z,r}]^T:[0,\infty) \to \mathbb{R}^3$, $\psi_r:[0,\infty) \to \mathbb{R}$ for the position and yaw angles with prescribed performance;
$\boldsymbol{p}_r$ and $\psi_r$ are assumed smooth functions of time with bounded first and second derivatives.  
Prescribed performance control, as described in Section~\ref{sec:preliminaries}, dictates
that the tracking error signal evolves strictly within a funnel
defined by prescribed functions of time, thus achieving desired
performance specifications, such as maximum overshoot, convergence speed, and maximum steady-state error. However,
notice that the UAV model \eqref{eq:system} is underactuated, and hence the original PPC methodology cannot be
directly applied. Consequently, we adapt the PPC
methodology to achieve trajectory tracking with prescribed
performance for the position and yaw-angle variables. More specifically, the control objective is to guarantee that the errors
\begin{subequations} \label{eq:position and yaw error}
\begin{align}
    \boldsymbol{e}_p &= \begin{bmatrix}
        e_{p_x} \\ e_{p_y} \\ e_{p_z}
    \end{bmatrix} = \boldsymbol{p} - \boldsymbol{p}_r \label{eq:position error} \\
    e_\psi &= \psi - \psi_r  \label{eq:yaw error}
\end{align}
\end{subequations}
evolve strictly within a funnel dictated by the corresponding exponential performance functions $\rho_{p_x}(t)$, $\rho_{p_y}(t)$, $\rho_{p_z}(t)$, $\rho_\psi(t)$, which is formulated as 
\begin{subequations} \label{eq:ppc objective}
\begin{align}
    |e_{p_i}(t)| < & \rho_{p_i}(t), \ i\in \{x,y,z\} \\
    |e_\psi(t) | < & \rho_\psi(t) 
\end{align}
\end{subequations}
for all $t \geq 0$, given the initial funnel compliance $|e_{p_i}(0)| < \rho_{p_i}(0)$, for $i\in \{x,y,z\}$, and $|e_\psi(0)| < \rho_\psi(0)$. The adopted exponentially-decaying performance functions are 
$\rho_{p_i}(t) = (\rho_{p_i,0} - \rho_{p_i,\infty} )\exp(-l_{p_i}t) + \rho_{p_i,\infty}$, $i\in \{x,y,z\}$, and 
$\rho_{\psi}(t) = (\rho_{\psi,0} - \rho_{\psi,\infty} )\exp(-l_{\psi}t) + \rho_{\psi,\infty}$. 

\section{MAIN RESULTS}\label{sec:main_results}
\subsection{Preliminaries}
Since the UAV system is underactuated, the idea is to take advantage of the specific control inputs to control the vertical velocity $v_z$ and angular velocity $\boldsymbol{\omega}$, but also introduce virtual control on the horizontal velocities $\boldsymbol{v}_{xy} = [v_x,v_y]^T$ such that the given reference is tracked. Let us first rewrite the system dynamics in a control suitable form. \\
Let 
\begin{equation*}
    \boldsymbol{R}_{z,\psi} = \begin{bmatrix}
\boldsymbol{R}_\psi & 0 \\
0 & 1
\end{bmatrix}
\end{equation*}
and the factorization 
\begin{equation*}
\boldsymbol{R}_{IB}  \boldsymbol{F}_{T} = \boldsymbol{R}_{z,\psi} \boldsymbol{R}_{y,\theta}\boldsymbol{R}_{x,\phi}\begin{bmatrix} 0 \\ 0 \\ F_z \end{bmatrix}= \begin{bmatrix}
\boldsymbol{R}_\psi  \boldsymbol{T}_{\phi\theta} F_z \\
 c_\theta c_\phi F_z
\end{bmatrix} 
\end{equation*}
where
\begin{equation}
    \boldsymbol{T}_{\phi\theta} = \begin{bmatrix}   s_\theta  c_\phi \\ - s_ \phi \end{bmatrix}, \label{eq:T_phitheta}
\end{equation}
and $\boldsymbol{R}_\psi \in SO(2)$. Then, the velocity dynamics $\boldsymbol{v} = [\boldsymbol{v}_{xy}^T, v_z]^T$ can be written as
\begin{subequations} \label{eq:v xyz dot}
\begin{align}
    \dot{\boldsymbol{v}}_{xy} &= \frac{1}{m}\left (\boldsymbol{R}_\psi\boldsymbol{T}_{\phi\theta}F_z + \boldsymbol{F}_{d,xy} \right ) , \\
    \dot{v}_z &= \frac{1}{m}\left (   c_\theta c_\phi F_z   + F_{d,z} \right ) - g.
\end{align}
\end{subequations}
Furthermore, let us define the matrices 
\begin{align*}
    \boldsymbol{J}_{\phi\theta} =& \begin{bmatrix} 
    - s_\theta  s_ \phi &  c_ \theta  c_ \phi \\ 
    -  c_ \phi & 0 
    \end{bmatrix} \\
    \boldsymbol{R}_{\phi\theta} =& \begin{bmatrix} 
    \frac{ c_\psi}{ c_\theta} & \frac{ s_\psi}{ c_\theta} \\
     - s_\psi &  c_\psi  \end{bmatrix},
\end{align*}
that satisfy $\dot{\boldsymbol{T}}_{\phi\theta} = \boldsymbol{J}_{\phi\theta} \begin{bmatrix} \dot{\phi} \\ \dot\theta \end{bmatrix}$, and 
\begin{equation*}
    \boldsymbol{R}_T = \begin{bmatrix} 
    \boldsymbol{R}_{\phi\theta} & \boldsymbol{0}_2 \\ 
      c_\psi  t_\theta \ \   s_\psi  t_ \theta  & 1
  \end{bmatrix}
\end{equation*}
and will be used in the sequel. Note that $\boldsymbol{R}_{\phi\theta}$ is well-defined and invertible for $|\theta| < \frac{\pi}{2}$, while $\boldsymbol{J}_{\phi\theta}$ is invertible for $|\phi| < \frac{\pi}{2}$ and $|\theta| < \frac{\pi}{2}$. This is a commonly used assumption for UAV systems \cite{huang2010adaptive,madani2006backstepping,raffo2009underactuated} and we adopt it in this paper: 
\begin{ass} \label{ass:singularity}
 The roll and pitch angles satisfy $|\phi(t)| \leq \bar{\pi}$, $|\theta(t)| \leq \bar{\pi}$, for all $t \geq 0$ and some $\bar{\pi} < \frac{\pi}{2}$.
\end{ass}

\subsection{Control Design}
We describe now the proposed control-design procedure. 
\subsubsection{PPC on position error}
We first define the normalized position error 
\begin{equation} \label{eq:ksi p}
    \boldsymbol{\xi}_{p} = \begin{bmatrix}
        \xi_{p_x} \\ \xi_{p_y} \\ \xi_{p_z}
    \end{bmatrix} = \boldsymbol{\rho}_p(t)^{-1} \boldsymbol{e}_{p}  
\end{equation}
where $\boldsymbol{\rho}_p = \textup{diag}\{ [\rho_{p_x},\rho_{p_y},\rho_{p_z}] \} \in \mathbb{R}^{3 \times 3}$. 
Next, we define the transformation 
\begin{align} \label{eq:epsilon p}
    \boldsymbol{\varepsilon}_p =& \mathrm{T}(\boldsymbol{\xi}_{p})
\end{align}
where $\mathrm{T}$ is given by \eqref{eq:transf_vec} and we design the reference velocity signal 
\begin{equation} \label{eq:v r}
    \boldsymbol{v}_r = 
    \begin{bmatrix}
    \boldsymbol{v}_{xy,r} \\ v_{z,r}
    \end{bmatrix}
    = - k_p \boldsymbol{\rho}_p^{-1} \boldsymbol{r}_p \boldsymbol{\varepsilon}_p
\end{equation}
where $\boldsymbol{r}_p = \frac{\mathrm{d} \mathrm{T}(\boldsymbol{\xi}_p)}{\mathrm{d} \boldsymbol{\xi}_p} = 
\textup{diag}\left\{ \frac{1}{1-\xi_{p_x}^2}, \frac{1}{1-\xi_{p_y}^2}, \frac{1}{1-\xi_{p_z}^2} \right\}$ and $k_p$ is a positive control gain. 

\subsubsection{PPC on velocity error}
Following a backstepping-like procedure, we define the error  
\begin{equation} \label{eq:e_v}
    \boldsymbol{e}_v = \begin{bmatrix}
    \boldsymbol{e}_{v_{xy}} \\ e_{v_z}
    \end{bmatrix} =  
    \begin{bmatrix}
    {e}_{v_x} \\ {e}_{v_y} \\ e_{v_z}
    \end{bmatrix} =
    \boldsymbol{v} - \boldsymbol{v}_{r} =
    \begin{bmatrix}
    \boldsymbol{v}_{xy} \\ v_{z}
    \end{bmatrix} 
    - 
    \begin{bmatrix}
    \boldsymbol{v}_{xy,r} \\ v_{z,r}
    \end{bmatrix}
\end{equation}
Next, we introduce the corresponding exponential performance functions $\rho_{v_i}(t) = (\rho_{v_i,0} - \rho_{v_i,\infty})\exp(-l_{v_i}t) + \rho_{v_i,\infty}$, such that $\rho_{v_i}(0) = \rho_{v_i,0} > |e_{v_i}(0)|$, for $i\in\{x,y,z\}$,  
which leads to the normalized error
\begin{equation} \label{eq:ksi v}
    \boldsymbol{\xi}_{v} = \begin{bmatrix}
    \boldsymbol{\xi}_{v_{xy}} \\ \xi_{v_z}
    \end{bmatrix} =    \begin{bmatrix}
    \xi_{v_x} \\ \xi_{v_y} \\ \xi_{v_z}
    \end{bmatrix} = \boldsymbol{\rho}_v(t)^{-1} \boldsymbol{e}_{v}, 
\end{equation}
with $\boldsymbol{\rho}_v = \textup{diag}\{ \rho_{v_x}, \rho_{v_y}, \rho_{v_z} \}$. 
Next, we define the transformation 
\begin{equation} \label{eq:epsilon v}
    \boldsymbol{\varepsilon}_v = 
    \begin{bmatrix}
        \boldsymbol{\varepsilon}_{v_{xy}} \\
        \varepsilon_{v_z}
    \end{bmatrix}
    =\mathrm{T}(\boldsymbol{\xi}_{v})
\end{equation}
and set the control input $F_z$ as 
\begin{equation} \label{eq:F z}
    F_z = -{k_{v_z}} \rho_{v_z}^{-1} r_{v_z} 
    \varepsilon_{v_z}
\end{equation}
where ${r}_{v_z} = \frac{\mathrm{d} \mathrm{T}({\xi}_{v_z})}{\mathrm{d} {\xi}_{v_z}} = 
 \frac{1}{1-\xi_{v_z}^2}$ and $k_{v_z}$ is a positive control gain. 
Moreover, we define the reference signal for $\boldsymbol{T}_{\phi \theta}$, defined in~\eqref{eq:T_phitheta}, as 
\begin{equation} \label{eq:T phi theta r}
    \boldsymbol{T}_{\phi \theta, r} = 
    \begin{bmatrix}
    T_{\phi\theta_1,r} \\
    T_{\phi\theta_2,r}
    \end{bmatrix}
    = 
    - k_{v_{xy}} \frac{\boldsymbol{R}_{\psi}^T\boldsymbol{\rho}_{v_{xy}}^{-1}\boldsymbol{r}_{v_{xy}}\boldsymbol{\varepsilon}_{v_{xy}}}{F_z},
\end{equation}
where $k_{v_{xy}}$ is a positive control gain, $\boldsymbol{\rho}_{v_{xy}} = \textup{diag}\{ \rho_{v_x}, \rho_{v_y} \}$, 
and $\boldsymbol{r}_{v_{xy}} = \frac{\mathrm{d} \mathrm{T}(\boldsymbol{\xi}_{v_{xy}})}{\mathrm{d} \boldsymbol{\xi}_{v_{xy}}} = 
\textup{diag}\left\{ \frac{1}{1-\xi_{v_x}^2}, \frac{1}{1-\xi_{v_y}^2} \right\}$.

\subsubsection{PPC on angular errors}:
The reference signal $\boldsymbol{T}_{\phi \theta, r}$ implicitly assigns reference values for the angles $\phi$, $\theta$. Moreover, given the reference $\psi_r$,    
we define the respective errors 
\begin{subequations} \label{eq:e phi theta}
\begin{align} 
    \boldsymbol{e}_{\phi \theta} &=\begin{bmatrix}
        e_{\phi \theta_1} \\ e_{\phi \theta_2}
    \end{bmatrix} 
    = \boldsymbol{T}_{\phi \theta} - \boldsymbol{T}_{\phi \theta, r} \\
    e_\psi &= \psi - \psi_r
\end{align}
\end{subequations}
By further introducing exponential performance functions $\rho_i(t) = (\rho_{i,0} - \rho_{i,\infty})\exp(-l_i t) +\rho_{i,\infty}$, such that $\rho_i(0) = \rho_{i,0} > |e_i(0)|$, for
$i \in \{\phi\theta_1, \phi\theta_2, \psi\}$, we define the normalized errors 
\begin{subequations} \label{eq:xi phi theta}
\begin{align} 
    \boldsymbol{\xi}_{\phi\theta} =&
    \begin{bmatrix}
        \xi_{\phi\theta_1} \\
        \xi_{\phi\theta_2}
    \end{bmatrix} =
    \boldsymbol{\rho}_{\phi\theta}(t)^{-1} \boldsymbol{e}_{\phi \theta} \\
    {\xi}_{\psi} =& \rho_\psi(t)^{-1} e_\psi 
\end{align}
\end{subequations}
where $\boldsymbol{\rho}_{\phi\theta} = \textup{diag}\{ \rho_{\phi\theta_1}, \rho_{\phi\theta_2} \}$,
and the transformations 
\begin{subequations} \label{eq:epsilon phi theta}
\begin{align} 
    \boldsymbol{\varepsilon}_{\phi \theta} &= \mathrm{T}( \boldsymbol{\xi}_{\phi\theta}) \\
    \varepsilon_\psi &= \mathrm{T}( {\xi}_{\psi})
\end{align}
\end{subequations}
In order to stabilize the aforementioned errors, we design reference signals for the angular velocities 
\begin{align} \label{eq:omega r}
   \boldsymbol{\omega}_r = \begin{bmatrix}
       \boldsymbol{\omega}_{\phi \theta, r} \\ 
       \omega_{\psi,r}
   \end{bmatrix} 
   & = \begin{bmatrix}
       {\omega}_{\phi,r} \\ {\omega}_{\theta, r} \\ 
       \omega_{\psi,r}
   \end{bmatrix} 
    \notag \\
    & \hspace{-3mm} =
   - \begin{bmatrix}
   k_{\phi\theta}
    \boldsymbol{R}_{\phi\theta}^{-1}
    \boldsymbol{J}_{\phi\theta}^{-1}
    \boldsymbol{\rho}_{\phi\theta}^{-1}\boldsymbol{r}_{\phi\theta}\boldsymbol{\varepsilon}_{\phi\theta}  \\
     k_\psi \rho_\psi^{-1} r_\psi \varepsilon_\psi +   \omega_\phi c_\psi t_\theta + \omega_\theta s_\psi t_\theta  
     \end{bmatrix}
\end{align}
where $k_{\phi\theta}$ and $k_\psi$ are positive control gains, and $\boldsymbol{r}_{\phi\theta} = \frac{\mathrm{d} \mathrm{T}(\boldsymbol{\xi}_{\phi\theta})}{\mathrm{d} \boldsymbol{\xi}_{\phi\theta}} = 
\textup{diag}\left\{ \frac{1}{1-\xi_{\phi\theta_1}^2}, \frac{1}{1-\xi_{\phi\theta_2}^2} \right\}$. Note that $\boldsymbol{R}_{\phi\theta}^{-1}$ and
    $\boldsymbol{J}_{\phi\theta}^{-1}$ are well-defined due to Assumption \ref{ass:singularity}.

\subsubsection{PPC on angular velocity errors}:
The final step is the design of the control inputs $\boldsymbol{\tau}$ for tracking of the reference angular velocities designed in the previous step. To that end, we define first the angular velocity errors 
\begin{align} \label{eq:e omega}
    \boldsymbol{e}_\omega = \begin{bmatrix}
        \boldsymbol{e}_{\omega_{\phi \theta}} \\ e_{\omega_\psi}
    \end{bmatrix} = 
    \boldsymbol{\omega} - \boldsymbol{\omega}_r = \begin{bmatrix}
        \boldsymbol{\omega}_{\phi \theta} \\ 
        \omega_\psi 
    \end{bmatrix}
    - 
    \begin{bmatrix}
        \boldsymbol{\omega}_{\phi \theta,r} \\ 
        \omega_{\psi,r} 
    \end{bmatrix}
\end{align}
We further introduce exponential performance functions $\rho_{\omega_i}(t) = (\rho_{\omega_i,0} - \rho_{\omega_i,\infty})\exp(-l_{\omega_i}t) +\rho_{\omega_i,\infty}$,  such that $\rho_{\omega_i}(0) = \rho_{\omega_i,0} > |e_{\omega_i}(0)|$, for $i\in \{\phi,\theta,\psi\}$, to impose predefined performance on the error $\boldsymbol{e}_\omega$, and define the respective normalized error
\begin{align} \label{eq:ksi omega}
    \boldsymbol{\xi}_\omega = \begin{bmatrix}
        \xi_{\omega_\phi} \\
        \xi_{\omega_\theta} \\
        \xi_{\omega_\psi}
    \end{bmatrix} 
    = \boldsymbol{\rho}_\omega(t)^{-1} \boldsymbol{e}_\omega 
\end{align}
Finally, we define the transformation 
\begin{align} \label{eq:epsilon omega}
    \boldsymbol{\varepsilon}_\omega = \mathrm{T}(\boldsymbol{\xi}_\omega)
\end{align}
and design the control input 
\begin{align} \label{eq:tau}
    \boldsymbol{\tau} = -k_\omega \boldsymbol{\rho}_\omega^{-1} \boldsymbol{r}_\omega \boldsymbol{\varepsilon}_\omega 
\end{align}
where $\boldsymbol{r}_\omega = \frac{\mathrm{d} \mathrm{T}(\boldsymbol{\xi}_{\omega})}{\mathrm{d} \boldsymbol{\xi}_{\omega}} = \textup{diag}\left\{  \frac{1}{1-\xi_{\omega_\phi}^2} ,\frac{1}{1-\xi_{\omega_\theta}^2},\frac{1}{1-\xi_{\omega_\psi}^2} \right\}$, $\boldsymbol{\rho}_\omega~=~\textup{diag}\{ \rho_{\omega_\phi},\rho_{\omega_\theta},\rho_{\omega_\psi} \}$, and $k_\omega$ is a positive gain.

\subsection{Stability Analysis}

We provide the stability analysis of the proposed control protocol in the next theorem:
\begin{thm} \label{thm:main}
Consider the UAV dynamics \eqref{eq:system} under the proposed control scheme \eqref{eq:ksi p}-\eqref{eq:tau} and Assumption \ref{ass:singularity}. Further assume that
\begin{subequations} \label{ass:theorem assumptions}
\begin{align}
   & F_z(t) \neq 0 \label{ass:Fz} \\
   & \frac{k_{v_{xy}}}{k_{v_z}} > \frac{\max_{i\in \{\phi\theta_1,\phi\theta_2\}} \{\rho_{i,0}\} }{ 4\cos(\bar{\pi})^2} \label{ass:gains} \\
   & |T_{i,r}(t)| \leq \rho_{i}(t) + 1, i \in \{\phi\theta_1,\phi\theta_2\} \label{ass:T}
\end{align}
\end{subequations}
for $t\geq 0$, where $\bar{\pi}$ is defined in Assumption \ref{ass:singularity}.  Then it holds that  
\begin{align*}
    |e_{p_i}(t)| < & \rho_{p_i}(t), \ i\in \{x,y,z\} \\
    |e_\psi(t) | < & \rho_\psi(t) 
\end{align*}
and all closed-loop signals are bounded, for all $t\geq 0$.
\end{thm}

\begin{rem}
Condition \eqref{ass:Fz} is needed for the boundedness of the intermediate signal \eqref{eq:T phi theta r}. Intuitively, it requires the UAV thrust, set in \eqref{eq:F z}, to be always positive and compensate for the gravitational force $\boldsymbol{g}$. Note that this is a condition encountered in the related literature (e.g., \cite{cunha2009nonlinear}).  
One can guarantee \eqref{ass:Fz} by adding an integrator in  \eqref{eq:F z} and adjust appropriately the performance function $\rho_{p_z}(t)$; for more details, we refer the reader to \cite{verginis2022robust}. 
The parameter-related condition \eqref{ass:gains}, which is needed for the correctness of Theorem \ref{thm:main}, essentially imposes restrictions on the angles $\theta, \phi$ and the error $\boldsymbol{e}_{\phi\theta}$; such error must be small enough, as dictated by the initial performance values $\rho_{\phi\theta_1,0}$, $\rho_{\phi\theta_2,0}$ in the nominator of the right-hand-side of \eqref{ass:gains}, and the angles $\theta$, $\phi$ themselves must be close to zero, maximizing the denominator of the right-hand-side of \eqref{ass:gains}. Intuitively, the aforementioned constraints require small reference signals $\boldsymbol{T}_{\phi\theta,r}$, which translate to  slow reference trajectories $p_{x,r}(t)$ and $p_{y,r}(t)$. Similarly, \eqref{ass:T} is needed since the values of $\boldsymbol{T}_{\phi\theta}$ in \eqref{eq:T_phitheta} cannot exceed the value of 1. We can enforce such a condition by choosing slowly-converging performance functions $\rho_{\phi\theta_1}(t)$ and $\rho_{\phi\theta_2}(t)$ with large initial values.  \label{rem:remark1}   
\end{rem}

\begin{proof}
The proof proceeds in three steps. First, we show the existence of a local solution such that $\boldsymbol{\xi}_p(t)$, $\boldsymbol{\xi}_v(t)$, $\boldsymbol{\xi}_\omega(t)$ $\in (-1,1)^3$, $\xi_\psi(t) \in (-1,1)$, $\boldsymbol{\xi}_{\phi\theta}(t) \in (-1,1)^2$ for a time interval $t \in [0,\tau_{\max})$.  Next, we show that the proposed control scheme retains the aforementioned normalized signals in compact subsets of $(-1,1)$, which leads to $\tau_{\max} = \infty$ in the final step, thus completing the proof.

Towards the existence of a local solution, consider first the overall state vector $\boldsymbol{\chi} = [ \boldsymbol{p}^T, \boldsymbol{v}^T, \boldsymbol{\eta}^T, \boldsymbol{\omega}^T]^T \in \mathbb{X} = \mathbb{R}^{6} \times (-\frac{\pi}{2}, \frac{\pi}{2})^2 \times (-{\pi}, {\pi})  \times \mathbb{R}^3$ and let us define the open set:
\begin{align}
    \Omega =& \big\{ (\boldsymbol{\chi},t) \in \mathbb{X} \times [0,\infty) : \boldsymbol{\xi}_p \in (-1,1)^3, \boldsymbol{\xi}_v \in (-1,1)^3, \notag \\
    &\boldsymbol{\xi}_{\phi \theta} \in (-1,1)^2, \xi_\psi \in (-1,1), \boldsymbol{\xi}_\omega \in (-1,1)^3 \big\}.
\end{align}
Note that the choice of the performance functions at $t=0$ implies that $\boldsymbol{\xi}_p(0)$, $\boldsymbol{\xi}_v(0)$, $\boldsymbol{\xi}_\omega(0)$ $\in (-1,1)^3$, $\boldsymbol{\xi}_{\phi\theta}(0) \in (-1,1)^2$, and $\xi_\psi(0) \in (-1,1)$, implying that $\Omega$ is nonempty. By combining \eqref{eq:system}, \eqref{eq:F z}, and \eqref{eq:tau}, we obtain the closed-loop system dynamics $\dot{\chi} = f_\chi(\chi,t)$, where $f_\chi: \mathbb{X} \times [0,\tau_{\max})$ is a function continuous in $t$ and locally Lipschitz in $\chi$. Hence, the conditions of Theorems~2.1.1(i) and 2.13 of \cite{bressan2007introduction} are satisfied and we conclude that there exists a unique and local solution $\chi:[0,\tau_{\max}) \to \mathbb{X}$ such that $(\boldsymbol{\chi}(t),t) \in \Omega$ for $t \in [0,\tau_{\max})$. Therefore, it holds that 
\begin{subequations} \label{eq:ksi local bound}
\begin{align}
    &\boldsymbol{\xi}_p \in (-1,1)^3 \\
    &\boldsymbol{\xi}_v \in (-1,1)^3 \\
    &\boldsymbol{\xi}_{\phi \theta} \in (-1,1)^2 \\
    &\xi_\psi \in (-1,1) \\
    &\boldsymbol{\xi}_\omega \in (-1,1)^3
\end{align}
\end{subequations}
for all $t \in [0,\tau_{\max})$. We next proceed to show that the normalized errors in \eqref{eq:ksi local bound} remain in compact subsets of $(-1,1)$. Note that  \eqref{eq:ksi local bound} implies that that transformed errors $\boldsymbol{\varepsilon}_p$, $\boldsymbol{\varepsilon}_v$, $\boldsymbol{\varepsilon}_{\phi\theta}$, $\varepsilon_\psi$, $\boldsymbol{\varepsilon}_\omega$ are well-defined for $t \in [0,\tau_{\max})$. Consider now the candidate Lyapunov function: 
\begin{align}
    V_p = \frac{1}{2} \| \boldsymbol{\varepsilon}_p \|^2
\end{align}
Differentiating $V_p$ along the local solution $\chi(t)$ yields
\begin{align}
    \dot{V}_p = \boldsymbol{\varepsilon}_p^T \boldsymbol{r}_p \boldsymbol{\rho}_p^{-1}( \boldsymbol{v} - \dot{\boldsymbol{p}}_r - \dot{\boldsymbol{\rho}}_p \boldsymbol{\xi}_p) 
\end{align}
By using $\boldsymbol{v} = \boldsymbol{e}_v + \boldsymbol{v}_r$, \eqref{eq:v r}, the boundedness of $\dot{\boldsymbol{p}}_r$, $\dot{\boldsymbol{\rho}}_p$, and \eqref{eq:ksi local bound}, $\dot{V}_p$ becomes
\begin{align}
    \dot{V}_p \leq -k_p \| \boldsymbol{\rho}_p^{-1} \boldsymbol{r}_p \boldsymbol{\varepsilon}_p \|^2 + \| \boldsymbol{\rho}_p^{-1} \boldsymbol{r}_p \boldsymbol{\varepsilon}_p \| \bar{F}_p
\end{align}
where $\bar{F}_p$ is a constant, independent of $\tau_{\max}$, satisfying $\| \boldsymbol{\rho}_v \boldsymbol{\xi}_v - \dot{\boldsymbol{p}}_r - \dot{\boldsymbol{\rho}}_p \boldsymbol{\xi}_p)  \| \leq \bar{F}_p$, for all $t \in [0,\tau_{\max})$. Therefore, we conclude that $\dot{V}_p < 0$ when $\| \boldsymbol{\rho}_p^{-1} \boldsymbol{r}_p \boldsymbol{\varepsilon}_p \| > \frac{\bar{F}_p}{ k_p}$. In view of the definition of $\boldsymbol{r}_p$, we conclude that 
$\dot{V} < 0$ when $\| \boldsymbol{\varepsilon}_p \| > \frac{\bar{F}_p \max_{i\in\{x,y,z\}}\{\rho_{p_i,0}\}}{ k_p}$. Hence, by invoking Theorem 4.18 of \cite{Khalil_nonlinear}, we conclude that 
\begin{subequations} \label{eq:bound ksi epsilon p}
\begin{align} 
    \| \boldsymbol{\varepsilon}_p \| \leq \bar{\varepsilon}_p = \max\left\{ \|\boldsymbol{\varepsilon}_p(0)\|, \frac{\bar{F}_p \max_{i\in\{x,y,z\}}\{\rho_{p_i,0}\}}{ k_p} \right\}
\end{align}
for $t \in [0,\tau_{\max})$, and by employing the inverse of \eqref{eq:transformation}, we obtain
\begin{align}  
 |{\xi}_{p_i}(t)| \leq \bar{\xi}_p = \tanh{\bar{\varepsilon}_p} < 1
\end{align}
\end{subequations}
for $t \in [0,\tau_{\max})$ and $i\in \{x,y,z\}$. Therefore, we conclude the boundedness of $\boldsymbol{v}_r(t)$ and $\boldsymbol{v}(t) = \boldsymbol{e}_v(t) + \boldsymbol{v}_r(t) = \boldsymbol{\rho}_v(t) \boldsymbol{\xi}_v(t) + \boldsymbol{v}_r(t)$ for all $t \in [0,\tau_{\max})$. By differentiating $\boldsymbol{v}_r(t)$ and using  \eqref{eq:bound ksi epsilon p}, we further conclude the boundedness of $\dot{\boldsymbol{v}}_r(t)$ for all $t \in [0,\tau_{\max})$. 

We consider next the function $V_v = \frac{1}{2} \|\boldsymbol{\varepsilon}_v\|^2$, whose derivative, in view of \eqref{eq:system}, and \eqref{eq:v xyz dot},  yields 
\begin{align*}
    \dot{V}_v =& \boldsymbol{\varepsilon}_{v_{xy}}^T \boldsymbol{r}_{v_{xy}} \boldsymbol{\rho}_{v_{xy}}^{-1} 
    \bigg(\frac{1}{m}(\boldsymbol{R}_\psi \boldsymbol{T}_{\phi\theta} F_z + \boldsymbol{F}_{d,xy})  \\ 
    &- \dot{\boldsymbol{v}}_{xy,r} - \dot{\boldsymbol{\rho}}_{v_{xy}}\boldsymbol{\xi}_{v_{xy}} \bigg) \\
    &\hspace{-3mm} + \varepsilon_{v_z}^T r_{v_z}\rho_{v_z}^{-1} \left(\frac{1}{m}\left (   c_\theta c_\phi F_z   + F_{d,z} \right ) -g- \dot{v}_{z,r}-\dot{\rho}_{v_z}\xi_{v_z}  \right)
\end{align*}
for $t \in [0,\tau_{\max})$. 
By using $\boldsymbol{T}_{\phi\theta} = \boldsymbol{e}_{\phi\theta} + \boldsymbol{T}_{\phi\theta,r}$, \eqref{eq:T phi theta r}, \eqref{eq:F z}, the boundedness of $\dot{\boldsymbol{v}}_r$, $\dot{\boldsymbol{\rho}}_v$, and \eqref{eq:ksi local bound}, and the continuity and boundedness of $\boldsymbol{F}_d(\boldsymbol{\chi},\dot{\boldsymbol{\chi}},t)$ in $(\boldsymbol{\chi},\dot{\boldsymbol{\chi}})$ and $t$, respectively, we arrive at 
\begin{align*}
    \dot{V}_v \leq & - \frac{k_{v_{xy}} }{m} \|\boldsymbol{\rho}_{v_{xy}}^{-1} \boldsymbol{r}_{v_{xy}} \boldsymbol{\varepsilon}_{v_{xy}} \|^2 - \frac{k_{v_z} c_\theta c_\phi }{m} \| \rho_{v_z}^{-1} r_{v_z} \varepsilon_{v_z} \|^2 \\ 
    & + \frac{k_{v_z} \bar{\rho}_{\phi\theta} }{m}\|  \boldsymbol{\rho}_{v_{xy}}^{-1} \boldsymbol{r}_{v_{xy}} \boldsymbol{\varepsilon}_{v_{xy}} \| \| \rho_{v_z}^{-1} r_{v_z} \varepsilon_{v_z} \| \\
    & + \|\boldsymbol{\rho}_v^{-1} \boldsymbol{r}_v \boldsymbol{\varepsilon}_v \| \bar{F}_v   
\end{align*}
where $\bar{F}_v$ is a positive constant, independent of $\tau_{\max}$, satisfying $\| \boldsymbol{F}_d - \boldsymbol{g} - \dot{\boldsymbol{v}}_{r} - \dot{\boldsymbol{\rho}}_v \boldsymbol{\xi}_v \| \leq \bar{F}_v$, for $t \in [0,\tau_{\max})$, and $\bar{\rho}_{\phi\theta} = \max_{i \in \{\phi\theta_1,\phi\theta_2\} } \{ \rho_{i,0}\}$. 
Additionally, Assumption \ref{ass:singularity} implies that $c_\theta c_\phi \geq \bar{c} = \cos(\bar{\pi})^2 > 0$. 
Let now a constant $\alpha$ such that $k_{v_{xy}} > \frac{k_{v_z} \sqrt{\alpha}}{ 2 }$, $\cos(\bar{\pi})^2 > \frac{\bar{\rho}_{\phi\theta}}{2\sqrt{\alpha}}$. Note that such a constant exists due to \eqref{ass:gains}. By completing the squares, $\dot{V}_v$ becomes
\begin{align*}
    \dot{V}_v \leq -\kappa_v \|\boldsymbol{\rho}_v^{-1} \boldsymbol{r}_v \boldsymbol{\varepsilon}_v \|^2 + \|\boldsymbol{\rho}_v^{-1} \boldsymbol{r}_v \boldsymbol{\varepsilon}_v \| \bar{F}_v 
\end{align*}
for $t \in [0,\tau_{\max})$,
where $\kappa_v = \min\{\kappa_{v_{xy}}, \kappa_{v_z}\}$, and 
$\kappa_{v_{xy}} = \frac{k_{v_{xy}}}{m} - \frac{k_{v_z}\sqrt{\alpha}}{2m}$, $\kappa_{v_z} = k_{v_z}\left(\frac{\bar{c}}{m} - \frac{ \bar{\rho_{\phi\theta}}}{2m\sqrt{\alpha}} \right)$. Therefore, by following a similar procedure as with $V_p$, we conclude that 
\begin{subequations} \label{eq:bound ksi epsilon v}
\begin{align}
    \|\boldsymbol{\varepsilon}_v(t)\| \leq & \bar{\varepsilon}_v = \max\left\{\|\boldsymbol{\varepsilon}_v(0)\|, \frac{ \bar{F}_v \max_{i\in \{x,y,z\}} \{ \rho_{v_i,0} \}  }{ \kappa_v}  \right\} \\ 
    |\xi_{v_i}(t)| \leq &  \bar{\xi}_v = \tanh \bar{\varepsilon}_v < 1
\end{align}
\end{subequations}
for $t \in [0,\tau_{\max})$ and $i \in \{x,y,z\}$. Therefore, also in view of \eqref{ass:Fz}, we conclude the boundedness of $\boldsymbol{T}_{\phi\theta,r}$, $F_z$, $\boldsymbol{T}_{\phi\theta} = \boldsymbol{\rho}_{\phi\theta}(t) +\boldsymbol{T}_{\phi\theta,r}$, for $t \in [0,\tau_{\max})$. By differentiating $\boldsymbol{T}_{\phi\theta,r}$ and $F_z$ and using 
\eqref{eq:bound ksi epsilon v} and \eqref{ass:Fz}, we further conclude the boundedndess of  $\dot{\boldsymbol{T}}_{\phi\theta,r}(t)$ and $\dot{F}_z(t)$, for all $t \in [0,\tau_{\max})$.

Following a similar line of proof, we consider now the function $V_{\eta} = \frac{1}{2}\|\boldsymbol{\varepsilon}_{\phi\theta}\|^2 + \frac{1}{2} \varepsilon_\psi^2$, whose derivative, in view of \eqref{eq:system}, becomes 
\begin{align*}
    \dot{V}_{\eta} =&  \boldsymbol{\varepsilon}_{\phi\theta}^T
     \boldsymbol{r}_{\phi\theta} 
     \boldsymbol{\rho}_{\phi\theta}^{-1}
     \left(
     \boldsymbol{J}_{\phi\theta}
     \boldsymbol{R}_{\phi\theta}
     \boldsymbol{\omega}_{\phi\theta}  
     - \dot{\boldsymbol{T}}_{\phi\theta,r}
     - \dot{\boldsymbol{\rho}}_{\phi\theta}
     \boldsymbol{\xi}_{\phi\theta} 
     \right) \\ 
    &+ \varepsilon_{\psi}\rho_{\psi}^{-1}r_{\psi}( c_\psi t_\theta  \omega_\phi+ s_\psi t_\theta  \omega_\theta + \omega_\psi - \dot{\psi}_r - \dot{\rho}_\psi \xi_\psi)
\end{align*}
By using $\boldsymbol{\omega}_{\phi\theta} = \boldsymbol{e}_{\omega_{\phi\theta}} + \boldsymbol{\omega}_{\phi\theta,r}$, $\psi = \psi_r + e_\psi$, \eqref{eq:omega r}, \eqref{eq:ksi local bound}, and the continuity of $\boldsymbol{J}_{\phi\theta}$, Assumption \ref{ass:singularity}, and the boundedness of $\dot{\boldsymbol{T}}_{\phi\theta,r}$, $\dot{\boldsymbol{\rho}}_{\phi\theta}$, $\rho_\psi$, one obtains
\begin{align*}
    \dot{V}_{\eta} \leq & -k_{\phi\theta} \|\boldsymbol{\rho}_{\phi\theta}^{-1} \boldsymbol{r}_{\phi\theta} \boldsymbol{\varepsilon}_{\phi\theta} \|^2 + \|\boldsymbol{\rho}_{\phi\theta}^{-1} \boldsymbol{r}_{\phi\theta} \boldsymbol{\varepsilon}_{\phi\theta} \|\bar{F}_{\phi\theta} \\
     & - k_\psi (\rho_{\psi}^{-1}r_{\psi} \varepsilon_\psi)^2 + |\rho_{\psi}^{-1}r_{\psi}\varepsilon_\psi| \bar{F}_\psi \\
     \leq & -k_\eta\| \boldsymbol{\rho}_\eta^{-1} \boldsymbol{r}_\eta \boldsymbol{\varepsilon}_\eta \|^2 + \| \boldsymbol{\rho}_\eta^{-1} \boldsymbol{r}_\eta \boldsymbol{\varepsilon}_\eta \| \bar{F}_\eta   
\end{align*}
where $\bar{F}_{\phi\theta}$ and $\bar{F}_\psi$ are positive constants, independent of $\tau_{\max}$, satisfying $\| \boldsymbol{J}_{\phi\theta} \boldsymbol{R}_{\phi\theta} \boldsymbol{e}_{\omega_{\phi\theta}} - \dot{\boldsymbol{T}}_{\phi\theta,r} - \dot{\boldsymbol{\rho}}_{\phi\theta} \boldsymbol{\xi}_{\phi\theta}\| \leq  \bar{F}_{\phi\theta}$ and $| e_\psi - \dot{\psi}_r - \dot{\psi} \xi_\psi  | \leq \bar{F}_\psi$, for $t \in [0,\tau_{\max})$, 
and we further define $\boldsymbol{\rho}_\eta = \textup{diag}\{ \boldsymbol{\rho}_{\phi\theta}, \rho_\psi \}$, $\boldsymbol{r}_\eta = \textup{diag}\{ \boldsymbol{r}_{\phi\theta}, r_\psi \}$, $\boldsymbol{\varepsilon}_\eta = [\boldsymbol{\varepsilon}_{\phi\theta}^T, \varepsilon_\psi]^T$, $k_\eta = \min\{ k_{\phi\theta}, k_\psi \}$, and $\bar{F}_\eta = \max\{\bar{F}_\psi, \bar{F}_{\phi\theta}\}$. 
Therefore, it holds that $\dot{V}_{\eta} < 0$ when 
$\|\boldsymbol{\rho}_{\eta}^{-1} \boldsymbol{r}_{\eta} \boldsymbol{\varepsilon}_{\eta} \| > \frac{\bar{F}_{\eta}}{k_{\eta}}$, which,
similar to the previous steps, leads to
\begin{subequations} \label{eq:bound ksi epsilon eta}
\begin{align}
    \|\boldsymbol{\varepsilon}_\eta(t) \| \leq & \bar{\varepsilon}_\eta =  \max\left\{  \|\boldsymbol{\varepsilon}_\eta(0) \|, \frac{\bar{F}_{\eta} \max_{i \in {\phi,\theta,\psi} } \{ \rho_{i,0}  \} }{k_{\eta}} \right\} \\
    |{\xi}_i(t)| \leq & \bar{\xi}_\eta =  \tanh(\bar{\varepsilon}_\eta)
\end{align}
\end{subequations}
for $t \in [0,\tau_{\max})$ and $i\in\{\phi\theta_1,\phi\theta_2,\psi\}$. Therefore, in view of the boundedness of $\boldsymbol{e}_\omega(t)$, we conclude the boundedness of $\boldsymbol{\omega}_{\phi\theta,r}(t)$, ${\omega}_{\psi,r}(t)$ and hence of $\omega(t)$,  
for all $t \in [0,\tau_{\max})$. By using \eqref{eq:bound ksi epsilon eta}, we further conclude the boundedness of $\dot{\boldsymbol{\omega}}_{\phi\theta,r}(t)$ and  $\dot{\omega}_{\psi,r}(t)$ for all $t \in [0,\tau_{\max})$.

Finally, using a similar line of proof and considering the function $V_\omega = \frac{1}{2}\|\boldsymbol{e}_\omega\|^2$, we conclude that 
\begin{subequations} \label{eq:bound ksi epsilon omega}
\begin{align}
     \| \boldsymbol{\varepsilon}_\omega(t) \| &\leq \bar{\varepsilon}_\omega = \max\left\{ \| \boldsymbol{\varepsilon}_\omega(0)\|, \frac{\bar{F}_\omega \max_{i\in \{ \phi,\theta,\psi \} } \{ \rho_{\omega_i,0} \}   }{ k_\omega \lambda} \right\} \\
     |\xi_{\omega_i}(t)| &\leq  \tanh(\bar{\varepsilon}_\omega ) < 1,
\end{align}
\end{subequations}
for $t \in [0,\tau_{\max})$ and $i\in\{\phi,\theta,\psi\}$, where  $\lambda$ is the positive minimum eigenvalue of the positive definite inertia matrix $\boldsymbol{I}(\boldsymbol{\eta})$; $\bar{F}_\omega$ is a positive constant satisfying $\| \boldsymbol{\tau}_d -  \boldsymbol{I}^{-1} \boldsymbol{\omega}\times \boldsymbol{I} \boldsymbol{\omega}   - \dot{\boldsymbol{\omega}}_r -  \dot{\boldsymbol{\rho}}_\omega \boldsymbol{\xi}_\omega   \| \leq \bar{F}_\omega$, where we use the boundedness of $\boldsymbol{\xi}_\omega$ and $\boldsymbol{\omega}$ from \eqref{eq:ksi local bound}, the boundedness of $\dot{\boldsymbol{\omega}}_{\phi\theta,r}(t)$ from the previous step, and the boudnedness of $\boldsymbol{\tau}_d$ due to its continuity in $(\boldsymbol{\chi},\dot{\boldsymbol{\chi}})$ and boundedness in $t$.
Finally, \eqref{eq:bound ksi epsilon omega}  leads to the boundedness of $\boldsymbol{\tau}(t)$ for all $t \in [0,\tau_{\max})$. 

What remains to be shown is that $\tau_{\max} = \infty$. Towards that end, note that \eqref{eq:bound ksi epsilon p}, \eqref{eq:bound ksi epsilon v}, \eqref{eq:bound ksi epsilon eta}, and \eqref{eq:bound ksi epsilon omega} imply that $(\boldsymbol{\chi}(t),t)$ remain in a compact subset of $\Omega$, i.e., there exists a positive constant $\underline{d}$ such that $d_\mathcal{S}((\boldsymbol{\chi}(t),t), \partial \Omega) \geq \underline{d} > 0$, for all $t \in [0,\tau_{\max})$. Since all closed-loop signals have already been proven bounded, it holds that 
$\lim_{t \to \tau_{\max}^-} \left( \|\boldsymbol{\chi}(t)\| + d_\mathcal{S}((\boldsymbol{\chi}(t),t), \partial \Omega)^{-1} \right) \leq \bar{d}$, for some finite constant $\bar{d}$, and hence direct application of Theorem~2.1.4 of \cite{bressan2007introduction} dictates that $\tau_{\max} = \infty$, which concludes the proof. \end{proof}

\begin{rem}
From the aforementioned proof it can be deduced
that the proposed control scheme achieves its goals without
resorting to the need of rendering the ultimate bounds $\bar{\varepsilon}_p$, $\bar{\varepsilon}_v$,$\bar{\varepsilon}_\eta$, $\bar{\varepsilon}_\omega$ of the transformed errors arbitrarily small by adopting extreme values of the control gains $k_p$, $k_{v_{xy}}$, $k_{v_z}$, $k_{\phi\theta}$, $k_\psi$, and $k_\omega$; notice that \eqref{eq:bound ksi epsilon p}, \eqref{eq:bound ksi epsilon v}, \eqref{eq:bound ksi epsilon eta}, and \eqref{eq:bound ksi epsilon omega} hold no matter how large the finite bounds
$\bar{\varepsilon}_p$, $\bar{\varepsilon}_v$,$\bar{\varepsilon}_\eta$, $\bar{\varepsilon}_\omega$ are and regardless of the choice of the control
gains. 
In the same spirit, large uncertainties involved in
the nonlinear model \eqref{eq:system} can be compensated, as they affect
only the size of these bounds through $\bar{F}_v$ and $\bar{F}_\omega$, but leave
unaltered the achieved stability properties. Hence, the actual
performance given in \eqref{eq:ppc objective}, which is solely determined by the
designer-specified performance functions, becomes isolated against model uncertainties,
thus extending greatly the robustness of the proposed control
scheme. 
\end{rem}

\begin{rem}
It should be noted that the selection of the control
gains affects both the quality of evolution of the errors $\boldsymbol{e}_p$, $e_\psi$ within the corresponding performance envelopes as well
as the control input characteristics. Additionally, fine tuning
might be needed in real-time scenarios, to retain the required
control input signals within the feasible range that can be
implemented by the actuators. Similarly, the control input
constraints impose an upper bound on the required speed of
convergence of $\rho_{p_i}(t)$, $i\in\{x,y,z\}$,
$\rho_\psi(t)$, as obtained by the exponentials
$\exp(-l_{p_i}t)$, $i\in\{x,y,z\}$, and
$\exp(-l_\psi t)$, respectively.  Hence, the selection of
the control gains $k_p$, $k_{v_{xy}}$, $k_{v_z}$, $k_{\phi\theta}$, $k_\psi$, $k_\omega$ can have positive
influence on the overall closed loop system response.
More
specifically, notice that $\bar{F}_v$ and $\bar{F}_\omega$ provide implicit bounds
on $\bar{\varepsilon}_v$ and $\bar{\varepsilon}_\omega$, respectively. Therefore, invoking \eqref{eq:F z} and \eqref{eq:tau},  we can select the control gains such that $F_z$ and $\boldsymbol{\tau}$ are
retained within certain bounds. Nevertheless, the constants $\bar{F}_v$ and $\bar{F}_\omega$ involve the parameters of the model and the external
disturbances. Thus, an upper bound of the dynamic parameters
of the system as well as of the exogenous disturbances should
be given in order to extract any relations between the achieved
performance and the input constraints.
Finally, we stress that the scalar control gains $k_p$, $k_{v_{xy}}$, $k_{v_z}$, $k_{\phi\theta}$, $k_\psi$, $k_\omega$ can be replaced by diagonal matrices, adding more flexibility in the control design, without affecting the stability analysis.  
\end{rem}

\section{SIMULATION RESULTS}\label{sec:simulation_results}
\begin{figure}[b!]
      \centering
      \vspace{-0.5cm}
      \includegraphics[width=0.80\linewidth]{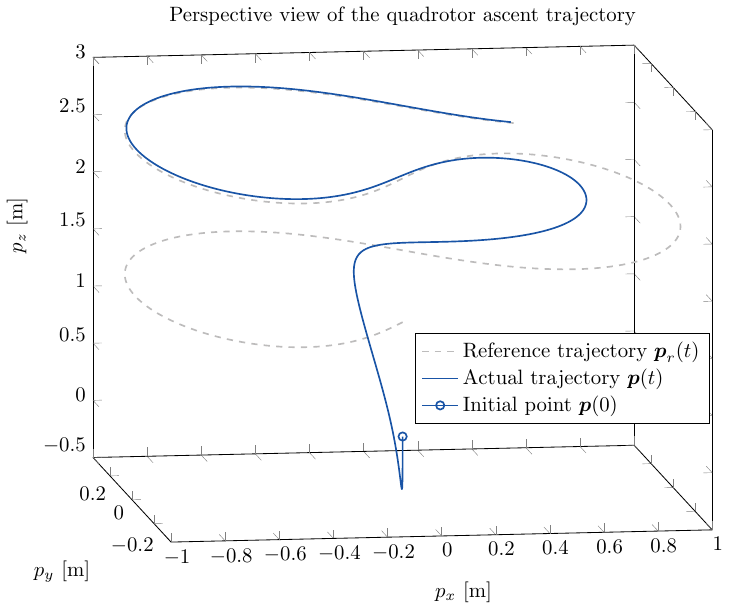}
      \caption{Perspective view of trajectory tracking. Note that initial drop in the actual trajectory occurs due to zero initial controlled thrust and no simulated surface beneath.}
      \label{fig:perspective}
\end{figure}
We evaluate the proposed control algorithm to the case of tracking reference trajectories $\boldsymbol{p}_r(t)=[p_{x,r}(t),p_{y,r}(t),p_{z,r}]^T$ in ascent and landing scenarios. To ensure that errors start inside of the funnel we choose the following prescribed performance functions: for the position $\rho_{p_i}(t)=(12-0.2)\exp{(-0.4t)}+0.2$, $i~=~\{x,y,z\}$, horizontal velocities $\rho_{v_i}(t)~=~(3-0.5)\exp{(-0.5t)}+0.5$, $i=\{x,y\}$ and the vertical velocity $\rho_{v_z}(t)~=~(5-0.2)\exp{(-1.5t)}+0.2$, angles $\rho_{\phi\theta_j}(t) =(0.5-0.25)\exp{(-0.5t)}+0.25$, $j=\{1,2\}$ and $\rho_\psi(t) = (0.4-0.05)\exp{(-0.1t)}+0.05$, finally, for angular velocities $\rho_{\omega_j}(t)=(0.3-0.1)\exp{(-0.5t)}+0.1$,$j=\{\phi,\theta,\psi\}$. 
The control gains are selected as $k_p = \textup{diag}\{1.25,1.25,12.5\}$, $k_{v_z} = 10$, $k_{v_{xy}} = \textup{diag}\{1,2\}$, $k_{\phi\theta} = \textup{diag}\{3,1.5\}$, $k_\psi = 1$, $k_\omega = 10 I_3$. The parameters are identical in both scenarios.
\begin{figure}[h!]
    \centering 
    \vspace{0.2cm}
      \includegraphics[width=\linewidth]{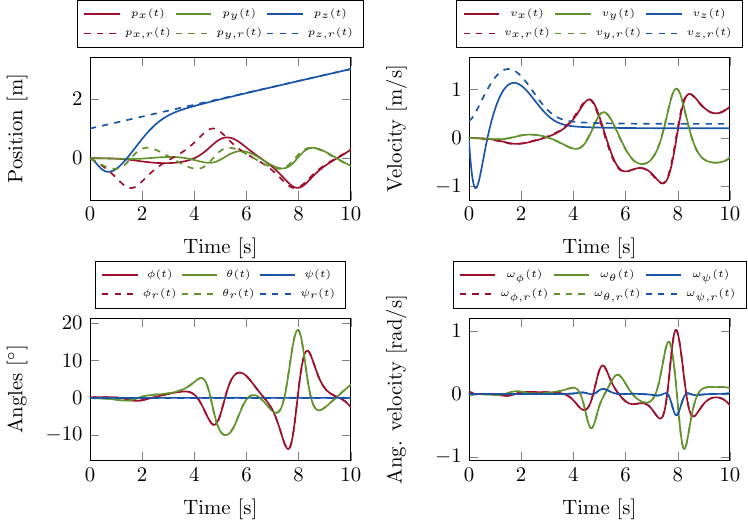}
      \caption{The evolution of state signals $\boldsymbol{p}(t), \boldsymbol{v}(t), \boldsymbol{\eta}(t), \boldsymbol{\omega}(t)$ compared to the given reference $\boldsymbol{p}_r(t), \psi_r(t)$ and the designed reference signals $\boldsymbol{v}_r(t),\boldsymbol{\omega}_r(t)$ as well as reference angles extracted from $\boldsymbol{T}_{\phi\theta,r}$.}
      \label{fig:real_vs_ref}
\end{figure}
\begin{figure}[h!]
      \centering
      \vspace{0.3cm}
      \includegraphics[width=0.95\linewidth]{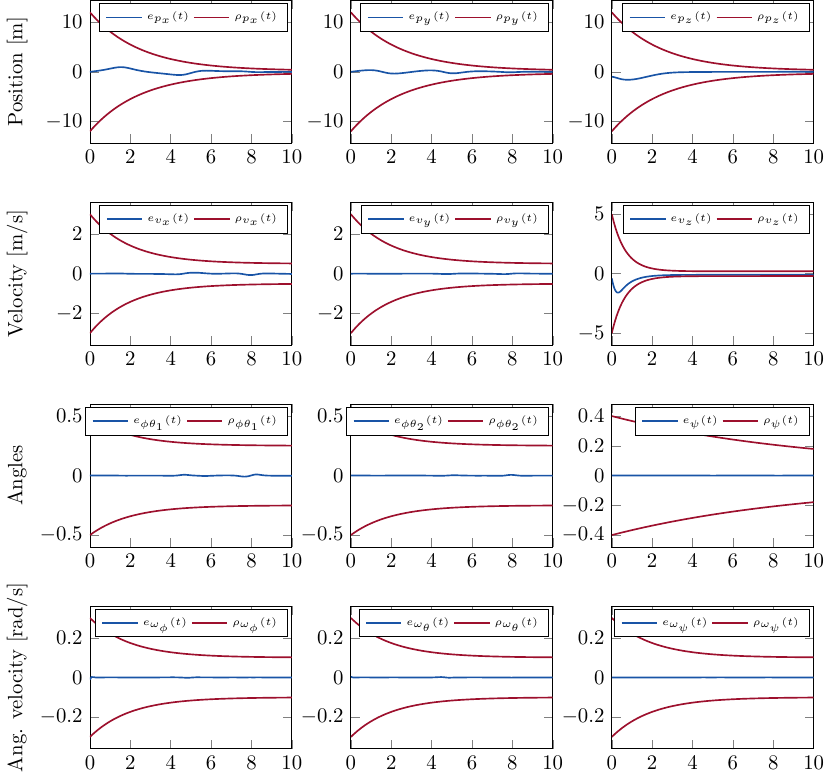}
      \caption{Error evolutions stay inside of the prescribed funnels during the experiment. Note that errors $\boldsymbol{e}_{\phi\theta} = [e_{\phi \theta_1}, e_{\phi \theta_2}]^T = \boldsymbol{T}_{\phi\theta} - \boldsymbol{T}_{\phi\theta,r}$ are dimensionless and $e_\psi$ is in radians.}

       \label{fig:errors_vs_funnels}
\end{figure}
\subsubsection{Ascent trajectory}
The reference ascent trajectory is constructed as a lemniscate ("$\infty$-shaped" trajectory) in the horizontal plane with the ramp function in the vertical direction and zero yaw reference, i.e. $p_{x,r}(t) = \cos(t)/(1+\sin^2(t)), p_{y,r}(t) = \sin(t)\cos(t)/(1+\sin^2(t)),  p_{z,r}(t) = 1+\frac{1}{5}t, \psi_r(t) = 0$ starting from the origin $\boldsymbol{p}(0) = [0,0,0]^T$. 
In Fig.~\ref{fig:perspective} we see how the big initial displacement from the reference signal is gradually reduced until the tracking error is eliminated. A comparison of all state signals during time against the given and designed references is provided in Fig.~\ref{fig:real_vs_ref}. Finally, in Fig.~\ref{fig:errors_vs_funnels} we observe that all error signals remain inside of the prescribed funnels during the simulation and that errors converge according to the specified performance functions.
\subsubsection{Landing scenario}
Quadrotor UAV is landing on a boat UGV moving according to a predefined trajectory $\boldsymbol{p}_b(t)=[p_{b,x}(t),p_{b,y}(t)]^T$ which is given as a solution of the following dynamical system with the control input $u(t)$ 
\begin{equation*}
\begin{matrix}
 \dot{p}_{b,x}(t)=\cos(\alpha(t)) \\
 \dot{p}_{b,y}(t)= \sin(\alpha(t))\\
 \dot{\alpha}(t)=u(t)
\end{matrix} \quad  u(t) = \left\{\begin{matrix} 
    -1& 0 \leq t \leq \frac{3\pi}{4}  \\ 
    1 & \frac{3\pi}{4} < t \leq \frac{9\pi}{4}  \\ 
    -1 & \frac{9\pi}{4} < t \leq \frac{11\pi}{4} \\ 
    0 & \frac{11\pi}{4} < t \leq 10 
\end{matrix}\right. 
\end{equation*}
It is assumed that the quadrotor is aware of the trajectory of the boat and the quadrotor reference is thus $\psi_r(t)=0$, $\boldsymbol{p}_r(t)=[p_{b,x}(t),p_{b,y}(t),p_{z,r}(t)]^T$, where $p_{z,r}(t) = z_d\left(1-\frac{1}{1+\exp(-(t-t_d))}\right)$, $z_d=5$ is the initial height and $t_d=5$ is a tuning parameter of the descent time.

\begin{figure}[h!]
      \centering
      \vspace{-0.2cm}
      \includegraphics[width=0.9\linewidth]{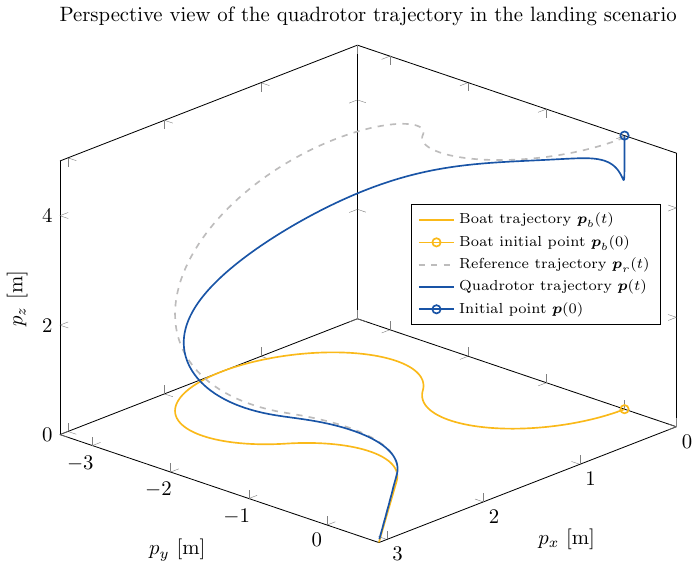}
      \caption{Perspective view of the landing scenario}
       \label{fig:perspective_landing}
\end{figure}
\begin{figure}[ht!]
      \centering 
      \vspace{0.25cm}
      \includegraphics[width=\linewidth]{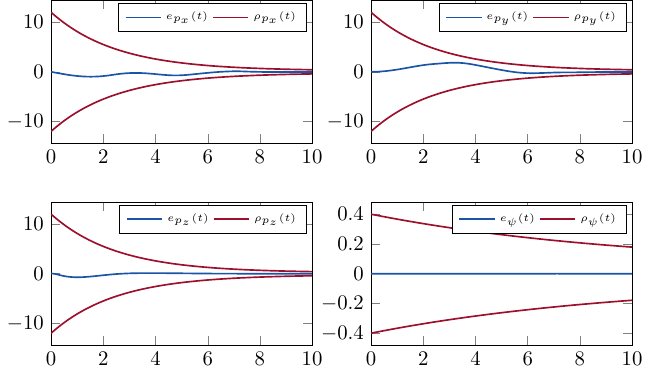}
      \caption{Error signals of the given references $\boldsymbol{p}_r(t),\psi_r(t)$ stay inside of the prescribed funnels during the experiment. The rest is omitted due to brevity.}
      \label{fig:errors_vs_funnels_landing}
\end{figure}

The perspective view of the landing is depicted in Fig.~\ref{fig:perspective_landing} and error signals evolution inside of funnels is shown in Fig.~\ref{fig:errors_vs_funnels_landing}. 

The errors between the references and states in the attitude subsystem in both cases are very small, thus achieving almost perfect tracking. The performance functions are chosen such that the roll and pitch angle errors through the transformation $\boldsymbol{T}_{\phi\theta}$ in \eqref{eq:T_phitheta}, are always less than $15^{\circ}$, approximately. This makes the control effort sufficiently aggressive while at the same time enables handling of stronger disturbances. The position subsystem errors are allowed to be quite big at the beginning of the transient and are sharply reduced to $0.2$m error bound. The most aggressive performance satisfaction is required on the vertical velocity $v_z$ to offset the gravity. This is in line with the discussion in Remark~\ref{rem:remark1}.
\vspace{-0.2cm}
\section{CONCLUSION}\label{sec:conclusion}
In this paper we presented approach in control design for the trajectory tracking of a quadrotor UAV using the Prescribed Performance Control methodology. Theoretical guarantees are established and controller is validated in simulation. Future work will include the deployment of the controller on the real quadrotor and testing it in experimental environments. Furthermore, we seek to extend the framework with trajectory generation and coordination in multi-agent scenarios.

\addtolength{\textheight}{-0.3cm}   






\bibliographystyle{unsrt}
\bibliography{root.bib}

\end{document}